\documentclass[twoside]{article}
\usepackage[accepted]{aistats2017}

\usepackage{microtype}
\usepackage{booktabs}
\usepackage{soul}
\usepackage{amsmath,amsfonts,amssymb,amsbsy}
\usepackage{url}
\usepackage{xcolor}
\usepackage{footnote}
\usepackage{graphicx} %
\usepackage{caption}
\usepackage{subcaption}
\usepackage{placeins} %
\usepackage[titletoc]{appendix} %
\usepackage{natbib} %
\usepackage{wrapfig}
\usepackage{listings}
\usepackage[colorlinks,linkcolor=black,citecolor=black,urlcolor=blue,filecolor=blue,backref=page]{hyperref}

\def\codesdir{codes/} %

\usepackage{ifthen}
\usepackage{tikz,pgfplots}
\usetikzlibrary{matrix}
\usetikzlibrary{calc}
\newlength{\figurewidth}
\newlength{\figureheight}

\def\figpdfdir{fig/} %
\def\figtikzdir{tikz/} %

\newcommand{\minput}[2][]{
\ifthenelse{\equal{#1}{pdf}}
	{ \includegraphics{\figpdfdir #2} }
	{ \tikzset{external/remake next} \tikzsetnextfilename{#2} \input{\figtikzdir #2} }
}

\usetikzlibrary{external}
\tikzset{external/system call={lualatex
	\tikzexternalcheckshellescape -halt-on-error -interaction=batchmode
	-jobname "\image" "\texsource"}}
\newcommand{\vc}[1] { \mathbf{#1} } %
\newcommand{\vs}[1] { \boldsymbol{#1} }
\newcommand{\tp}{\mathsf{T}}
\newcommand{\ti}[1] { \tilde{#1} } 
\newcommand{\mc}[1] { \mathcal{#1} } 
\newcommand{\tx}[1] { \text{#1} } 
\newcommand{\given} { \,|\, }

\newcommand{\mean}[2][] { \mathrm{E}_{#1} {\left[#2\right]} }
\newcommand{\var}[1] { \mathrm{Var} {\left[#1\right]} }

\newcommand{\Normal}[1] { \mathrm{N} {\left(#1\right)}  }
\newcommand{\halfNormal}[1] { \mathrm{N}^+ {\left(#1\right)}  }

\newcommand{\halfCauchy}[1] { \mathrm{C}^+ {\left(#1\right)} }
\newcommand{\Beta}[1] {\mathrm{Beta}{\left(#1\right)} }

\begin{document}

\runningtitle{Global shrinkage in the horseshoe prior}

\twocolumn[

\aistatstitle{On the Hyperprior Choice for the Global Shrinkage Parameter \\in the Horseshoe Prior}

\aistatsauthor{ Juho Piironen \And Aki Vehtari  }
\aistatsaddress{ {\tt juho.piironen@aalto.fi} \And {\tt aki.vehtari@aalto.fi} } 
\vspace{-1.6em}
\aistatsaddress{Helsinki Institute for Information Technology, HIIT \\ Department of Computer Science, Aalto University}

]

\begin{abstract}
The horseshoe prior has proven to be a noteworthy alternative for sparse Bayesian estimation, but as shown in this paper, the results can be sensitive to the prior choice for the global shrinkage hyperparameter.
We argue that the previous default choices are dubious due to their tendency to favor solutions with more unshrunk coefficients than we typically expect a priori.
This can lead to bad results if this parameter is not strongly identified by data.
We derive the relationship between the global parameter and the effective number of nonzeros in the coefficient vector, and show an easy and intuitive way of setting up the prior for the global parameter based on our prior beliefs about the number of nonzero coefficients in the model.
The results on real world data show that one can benefit greatly -- in terms of improved parameter estimates, prediction accuracy, and reduced computation time -- from transforming even a crude guess for the number of nonzero coefficients into the prior for the global parameter using our framework.
\end{abstract}

\section{INTRODUCTION}

With modern rich datasets, statistical models with a large number of parameters are nowadays commonplace in many application areas.
A typical example is a regression or classification problem with a large number of predictor variables.
In such problems, a careful formulation of the prior distribution  -- or regularization in the frequentist framework -- plays a key role.

Often it is reasonable to assume that only some of the model parameters $\vs \theta=(\theta_1,\dots,\theta_D)$ (such as the regression coefficients) are far from zero.
In the frequentist literature, these problems are typically handled by LASSO \citep{tibshirani1996} or one of its close cousins, such as the elastic net \citep{zou2005}. 
We focus on the probabilistic approach and carry out the full Bayesian inference on the problem.

Two prior choices dominate the Bayesian literature: two component discrete mixture priors known as the spike-and-slab \citep{mitchell1988,george1993}, and a variety of continuous shrinkage priors \citep[see e.g.,][and references therein]{polson2011}.
The spike-and-slab prior is intuitively appealing as it is equivalent to Bayesian model averaging (BMA) \citep{hoeting1999} over the variable combinations, and often has good performance in practice.
The disadvantages are that the results can be sensitive to prior choices (slab width and prior inclusion probability) and that the posterior inference can be computationally demanding with a large number of variables, due to the huge model space.
The inference could be sped up by analytical approximations using either expectation propagation (EP) \citep{hernandezlobato2010,hernandezlobato2015} or variational inference (VI) \citep{titsias2011}, but this comes at the cost of a substantial increase in the amount of analytical work needed to derive the equations separately for each model and a more complex implementation.

The continuous shrinkage priors on the other hand are easy to implement, provide convenient computation using generic sampling tools such as Stan \citep{stan_manual}, and can yield as good or better results.
A particularly interesting example is the horseshoe prior \citep{carvalho2009,carvalho2010}
\begin{align}
\begin{split}
	\theta_j \given \lambda_j,\tau &\sim \Normal{0,\lambda_j^2 \tau^2}\,, \\
	\lambda_j &\sim \halfCauchy{0,1}\,, \quad j=1,\dots,D,
\end{split}
\label{eq:hs_generic}
\end{align}
which has shown comparable performance to the spike-and-slab prior in a variety of examples where a sparsifying prior on the model parameters $\theta_j$ is desirable \citep{carvalho2009,carvalho2010,polson2011}.
The horseshoe is one of the so called global-local shrinkage priors, meaning that there is a global parameter $\tau$ that shrinks all the parameters towards zero, while the heavy-tailed half-Cauchy priors allow some parameters $\theta_j$ to escape the shrinkage (see Section~\ref{sec:horseshoe} for more thorough discussion).

So far there has been no consensus on how to carry out inference for the global hyperparameter $\tau$ which determines the overall sparsity in the parameter vector $\vs \theta$ and therefore has a large impact on the results.
For reasons discussed in Section~\ref{sec:full_bayes_vs_point_estimation}, we prefer full Bayesian inference.
For many interesting datasets, $\tau$ may not be well identified by data, and in such situations the hyperprior choice  $p(\tau)$ becomes crucial.
This is the question we address in this paper.

The novelty of the paper is summarized as follows.
We derive analytically the relationship between $\tau$ and $m_\tx{eff}$ -- the effective number of nonzero components in $\vs \theta$ (to be defined later) -- and show an easy and intuitive way of formulating the prior for $\tau$ based on our prior beliefs about the sparsity of $\vs \theta$.
We focus on regression and classification, but the methodology is applicable also to other generalized linear models and to other shrinkage priors than the horseshoe.
We argue that the previously proposed default priors are dubious based on the prior they impose on $m_\tx{eff}$, and that they yield good results only when $\tau$ (and therefore $m_\tx{eff}$) is strongly identified by the data.
Moreover, we show with several real world examples that in those cases where $\tau$ is only weakly identified by data, one can substantially improve inferences by transforming even a crude guess of the sparsity level into $p(\tau)$ using our method.

We first briefly review the key properties of the horseshoe prior in Section~\ref{sec:horseshoe}, and then proceed to discuss the prior choice $p(\tau)$ in Section~\ref{sec:global_parameter}.
The importance of the concept is illustrated in Section~\ref{sec:experiments} with real world data.
Section~\ref{sec:conclusion} concludes the paper by giving some recommendations on the prior choice based on the theoretical considerations and numerical experiments.

\section{THE HORSESHOE PRIOR FOR LINEAR REGRESSION}
\label{sec:horseshoe}

Consider the single output linear Gaussian regression model with several input variables, given by 
\begin{align}
\begin{split}
	y_i &= \vs \beta^\tp \vc x_i + \varepsilon_i, \quad \varepsilon_i \sim \Normal{0,\sigma^2}, \quad i=1,\dots,n \,, 
\end{split}
\label{eq:lgm}
\end{align}
where $\vc x$ is the $D$-dimensional vector of inputs, $\vs \beta$ contains the corresponding weights and $\sigma^2$ is the noise variance.
The horseshoe prior is set for the regression coefficients $\vs \beta = (\beta_1,\dots,\beta_D)$ 
\begin{align}
\begin{split}
	\beta_j \given \lambda_j,\tau &\sim \Normal{0,\lambda_j^2 \tau^2}, \\
	\lambda_j &\sim \halfCauchy{0,1} \,, \quad j=1,\dots,D.
\end{split}
\label{eq:hs}
\end{align}
If an intercept term $\beta_0$ is included in the model~\eqref{eq:lgm}, we give it a relatively flat prior, because there is usually no reason to shrink it towards zero.
As discussed in the introduction, the horseshoe prior has been shown to possess several desirable theoretical properties and good performance in practice \citep{carvalho2009,carvalho2010,polson2011,datta2013,vanDerPas2014}.
The intuition is the following: the global parameter $\tau$ pulls all the weights globally towards zero, while the thick half-Cauchy tails for the local scales $\lambda_j$ allow some of the weights to escape the shrinkage.
Different levels of sparsity can be accommodated by changing the value of $\tau$: with large $\tau$ all the variables have very diffuse priors with very little shrinkage towards zero, but letting $\tau \rightarrow 0$ will shrink all the weights $\beta_j$ to zero.

The above can be formulated more formally as follows.
Let $\vc X$ denote the $n$-by-$D$ matrix of observed inputs and $\vc y$ the observed targets.
The conditional posterior for the coefficients $\vs \beta$ given the hyperparameters and data $\mc D = (\vc X, \vc y)$ can be written as
\begin{align*}
	p(\vs \beta \given \vs \Lambda, \tau,\sigma^2,\mc D) &= \Normal{\vs \beta \given \vs {\bar \beta}, \vs \Sigma},\\
	\vs {\bar \beta} &= \tau^2 \vs \Lambda \left( \tau^2 \vs \Lambda + \sigma^2 (\vc X^\tp \vc X)^{-1} \right)^{-1} \vs {\hat \beta}, \\
	\vs \Sigma &= (\tau^{-2}\vs \Lambda^{-1} + \frac{1}{\sigma^2} \vc X^\tp \vc X)^{-1},
\end{align*}
where $\vs \Lambda = \tx{diag}(\lambda_1^2,\dots,\lambda_D^2)$ and $\vs {\hat \beta} = (\vc X^\tp \vc X)^{-1} \vc X^\tp \vc y$ is the maximum likelihood solution (assuming $(\vc X^\tp \vc X)^{-1}$ exists).
If the predictors are uncorrelated with zero mean and unit variance, then $\vc X^\tp \vc X \approx n \vc I$, and we can approximate
\begin{align}
	\bar \beta_j = (1-\kappa_j)\hat \beta_j,
\label{eq:beta_mean}
\end{align}
where
\begin{align}
	\kappa_j = \frac{1}{1+n \sigma^{-2}\tau^2\lambda_j^2}
\label{eq:shrinkage_factor}
\end{align}
is the {\it shrinkage factor} for coefficient $\beta_j$.
The shrinkage factor describes how much coefficient $\beta_j$ is shrunk towards zero from the maximum likelihood solution ($\kappa_j=1$ meaning complete shrinkage and $\kappa_j=0$ no shrinkage).
From~\eqref{eq:beta_mean} and~\eqref{eq:shrinkage_factor} it is easy to verify that $\vs {\bar \beta} \rightarrow 0$ as $\tau \rightarrow 0$, and $\vs {\bar \beta} \rightarrow \vs {\hat \beta}$ as $\tau \rightarrow \infty$.

\begin{figure}[t]
\centering
	\setlength{\figureheight}{0.15\textwidth} 
	\setlength{\figurewidth}{0.25\textwidth}
	\pgfplotsset{
	compat=newest,
	major tick length={0.03cm},
	x tick label style={font=\tiny},
	y tick label style={font=\tiny},
	ylabel style={text width=5em, rotate=-90, align=left},
	legend style={font=\tiny},
	} 
	\minput[pdf]{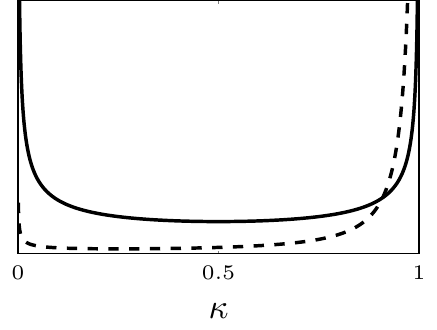}
	\caption{Density for the shrinkage factor~\eqref{eq:shrinkage_factor} for the horseshoe prior~\eqref{eq:hs} when $n \sigma^{-2}\tau^2 = 1$ (solid) and when $n \sigma^{-2}\tau^2 = 0.1$ (dashed).}
	\label{fig:kappa_prior}
\end{figure}

The result~\eqref{eq:shrinkage_factor} holds for any prior that can be written as a scale mixture of Gaussians like~\eqref{eq:hs}, regardless of the prior for $\lambda_j$.
The horseshoe employs independent half-Cauchy priors for all $\lambda_j$, and for this choice one can show that, for fixed $\tau$ and $\sigma$, the shrinkage factor~\eqref{eq:shrinkage_factor} follows the prior
\begin{align}
	p(\kappa_j\given \tau,\sigma) &= \frac{1}{\pi} \frac{\sigma^{-1}\tau \sqrt{n}}{(n\sigma^{-2}\tau^2-1)\kappa_j + 1}
								\frac{1}{\sqrt{\kappa_j} \sqrt{1-\kappa_j} }.
\label{eq:shrinkage_factor_prior}
\end{align}
When $n \sigma^{-2} \tau^2 = 1$, this reduces to $\Beta{\frac{1}{2},\frac{1}{2}}$ which looks like a horseshoe, see Figure~\ref{fig:kappa_prior}.
Thus, a priori, we expect to see both relevant ($\kappa=0$, no shrinkage) and irrelevant ($\kappa=1$, shrinkage) variables.
By changing the value of $\tau$, the prior for $\kappa$ places more mass either close to 0 or 1.
For instance, choosing $\tau$ so that $n \sigma^{-2} \tau^2 = 0.1$ favors complete shrinkage ($\kappa=1$) and thus we expect more variables to be shrunk a priori.
In the next section we discuss an intuitive way of designing a prior distribution for $\tau$ based on assumptions about the number of nonzero components in $\vs \beta$.

\section{THE GLOBAL SHRINKAGE PARAMETER}
\label{sec:global_parameter}

This section discusses the prior choice for the global hyperparameter $\tau$.
We begin with a short note on why we prefer full Bayesian inference for $\tau$ over point estimation, and then go on to discuss how we propose to set up the prior $p(\tau)$.

\subsection{Full Bayes versus Point Estimation}
\label{sec:full_bayes_vs_point_estimation}

In principle, one could use a plug-in estimate for $\tau$, obtained either by cross-validation or maximum marginal likelihood (sometimes referred to as ``empirical Bayes'').
The maximum marginal likelihood estimate has the drawback that it is always in danger of collapsing to $\hat \tau=0$ if the parameter vector happens to be very sparse.
Moreover, rather than being computationally convenient, this approach might actually complicate matters as the marginal likelihood is not analytically available for non-Gaussian likelihoods.
While cross-validation avoids the latter problem and possibly also the first one, it is computationally less efficient than the full Bayesian solution and fails to account for the posterior uncertainty.
For these reasons we recommend full Bayesian inference for $\tau$, and focus on how to specify the prior distribution.

\subsection{Earlier Approaches}
\label{sec:earlier_approaches}

\cite{carvalho2009} also recommend full Bayesian inference for $\tau$, and following \cite{gelman2006}, they propose prior
\begin{align}
	\tau \sim \halfCauchy{0,1},
\label{eq:tau_prior_hc1}
\end{align} 
whereas \cite{polson2011} recommend
\begin{align}
	\tau \given \sigma \sim \halfCauchy{0,\sigma^2}.
\label{eq:tau_prior_hcsigma}
\end{align}
Here $\halfCauchy{0,a^2}$ denotes the half-Cauchy distribution with location 0 and scale $a$.
If the target variable $y$ is scaled to have marginal variance of 1, unless the noise level $\sigma$ is very small, both of these priors typically lead to quite similar posteriors.
However, as we argue in Section~\ref{sec:meff_and_tau}, there is a theoretical justification for letting $\tau$ scale with $\sigma$.
The main motivation for using a half-Cauchy prior for $\tau$ is that it evaluates to a finite positive value at the origin, yielding a proper posterior and allowing even complete shrinkage $\tau \rightarrow 0$, while still having a thick tail which can accommodate a wide range of values.
For these reasons, $\halfCauchy{0,a^2}$ is a desirable choice when there are enough observations to let $\tau$ be identified by data. 
Still, we show that in several cases one can clearly benefit by choosing the scale $a$ in a more careful manner than simply $a=1$ or $a=\sigma$, because for most applications these choices place far to much mass for implausibly large values of $\tau$.
This point is discussed in Section~\ref{sec:meff_and_tau}.
Moreover, in some cases, $\tau$ can be so weakly identified by the data that one can obtain even better results by using a more informative and tighter prior, such as half-normal, in place of half-Cauchy (see the experiments in Section~\ref{sec:experiments}).

\cite{vanDerPas2014} study the optimal selection of $\tau$ in the model
\begin{align}
	y_i \sim \beta_i + \varepsilon_i, \quad \varepsilon_i \sim \Normal{0,\sigma^2}, \quad i=1,\dots,n.
\label{eq:model_simple}
\end{align}
They prove that in such a setting, the optimal value (up to a log factor) in terms of mean squared error and posterior contraction rates in comparison to the true $\vs \beta^*$ is 
\begin{align}
	\tau^* = \frac{p^*}{n},
\label{eq:vanderpas}
\end{align}
where $p^*$ denotes the number of nonzeros in the true coefficient vector $\vs \beta^*$ (assuming such exists).
Their proofs assume that $n,p^* \rightarrow \infty$ and $p^* = o(n)$.
Model~\eqref{eq:model_simple} corresponds to setting $\vc X = \vc I$ and $D=n$ in the usual regression model~\eqref{eq:lgm}.
It is unclear whether and how this result could be extended to a more general $\vc X$, and how one should utilize this result when $p^*$ is unknown (as it usually is in practice).
In the next section, we formulate our method of constructing the prior $p(\tau)$ based on prior beliefs about $p^*$, and show that if $p^*$ was known, our method would also give rise to result~\eqref{eq:vanderpas}, but is more generally applicable.

\subsection{Effective Number of Nonzero Coefficients}
\label{sec:meff_and_tau}

Consider the prior distribution for the shrinkage factor of the $j$th regression coefficient, Eq.~\eqref{eq:shrinkage_factor_prior}.
The mean and variance can be shown to be
\begin{align}
	\mean{\kappa_j \given \tau,\sigma} &= \frac{1}{1+\sigma^{-1}\tau\sqrt{n}}, 
	\label{eq:kappa_mean} \\
	\var{\kappa_j \given \tau,\sigma} &= \frac{\sigma^{-1} \tau \sqrt{n} }{ 2 (1 + \sigma^{-1}\tau \sqrt{n})^2}.
	\label{eq:kappa_var}
\end{align}
A given value for the global parameter $\tau$ can be understood intuitively via the prior distribution that it imposes on the effective number of coefficients distinguishable from zero (or effective number of nonzero coefficients, for short)
\begin{align}
	m_\tx{eff} = \sum_{j=1}^D (1-\kappa_j).
\label{eq:m_eff}
\end{align}
When the shrinkage factors $\kappa_j$ are close to 0 and 1 (as they typically are for the horseshoe prior), this quantity describes essentially how many active or unshrunk variables we have in the model.
It serves therefore as a useful indicator of the effective model size.

Using results~\eqref{eq:kappa_mean} and~\eqref{eq:kappa_var}, the mean and variance of $m_\tx{eff}$ given $\tau$ and $\sigma$ are given by
\begin{align}
	\mean{m_\tx{eff}\given\tau, \sigma} &= \frac{\sigma^{-1}\tau\sqrt{n} }{1+\sigma^{-1}\tau\sqrt{n}} D, \label{eq:meff_mean} \\
	\var{m_\tx{eff}\given\tau, \sigma} &= \frac{\sigma^{-1}\tau \sqrt{n} }{ 2 (1 + \sigma^{-1}\tau \sqrt{n})^2} D. \label{eq:meff_var}
\end{align}
The expression for the mean~\eqref{eq:meff_mean} is helpful.
First, from this expression it is evident that to keep our prior beliefs about $m_\tx{eff}$ consistent, $\tau$ must scale as $\sigma/\sqrt{n}$.
Priors that fail to do so, such as~\eqref{eq:tau_prior_hc1}, favor models of varying size depending on the noise level $\sigma$ and the number of data points $ n$. 
Second, if our prior guess for the number of relevant variables is $p_0$, it is reasonable to choose the prior so that most of the prior mass is located near the value
\begin{align}
	\tau_0 =  \frac{p_0}{D-p_0} \frac{\sigma}{\sqrt{n}},
\label{eq:tau0}
\end{align}
which is obtained by solving equation $\mean{m_\tx{eff}\given \tau,\sigma} = p_0$.
Note that this is typically quite far from $1$ or $\sigma$, which are used as scales for priors~\eqref{eq:tau_prior_hc1} and~\eqref{eq:tau_prior_hcsigma}.
For instance, if $D = 1000$ and $n = 200$, then prior guess $p_0=5$ gives about $\tau_0=3.6\cdot10^{-4} \sigma$.

To further develop the intuition about the connection between $\tau$ and $m_\tx{eff}$, it is helpful to visualize the prior imposed on $m_\tx{eff}$ for different prior choices for $\tau$.
This is most conveniently done by drawing samples for $m_\tx{eff}$\,; we first draw $\tau \sim p(\tau)$ and $\lambda_1,\dots,\lambda_D \sim \halfCauchy{0,1}$, then compute the shrinkage factors $\kappa_1,\dots,\kappa_D$ from ~\eqref{eq:shrinkage_factor}, and finally $m_\tx{eff}$ from~\eqref{eq:m_eff}. 

\begin{figure*}[t]
\centering
	\setlength{\figureheight}{0.3\textwidth}
	\setlength{\figurewidth}{0.85\textwidth}
	\pgfplotsset{
	compat=newest,
	major tick length={0.03cm},
	x tick label style={font=\tiny},
	y tick label style={font=\tiny},
	ylabel style={text width=5em, rotate=-90, align=left},
	legend style={font=\tiny},
	} 
	\minput[pdf]{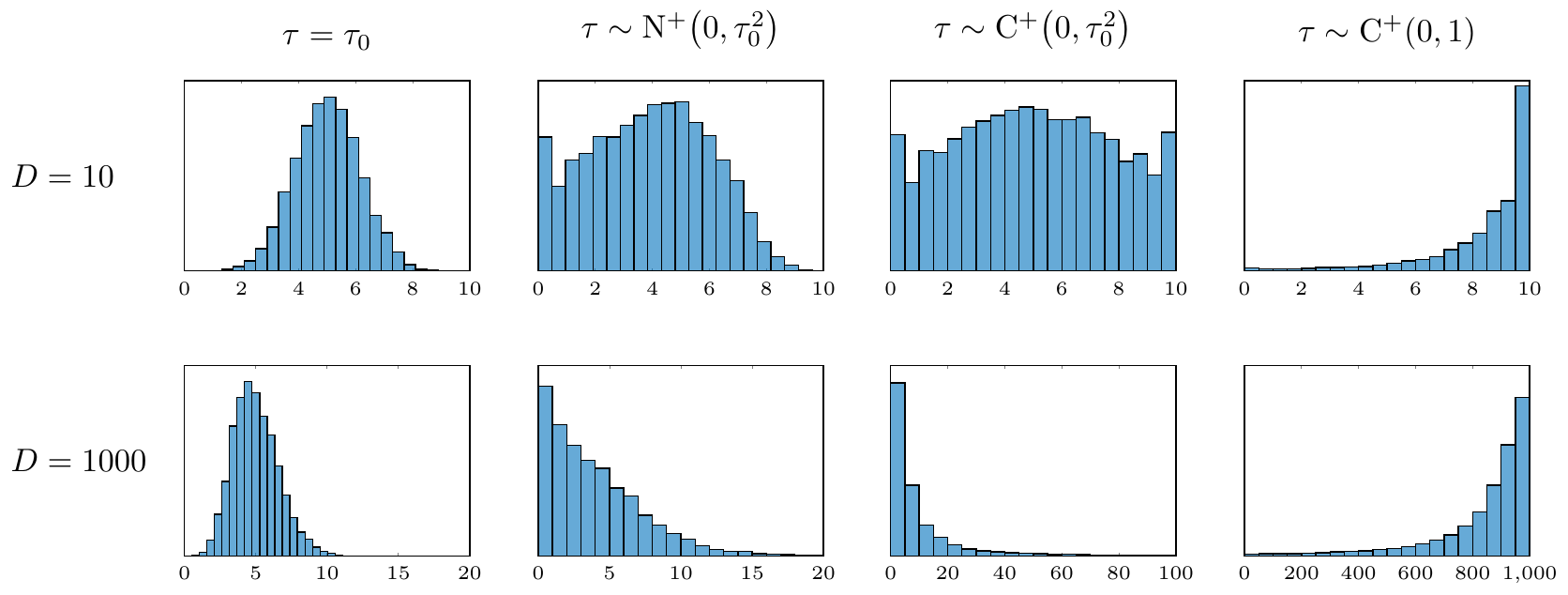}
	\caption{Histograms of prior draws for $m_\tx{eff}$ (effective number of nonzero regression coefficients, Eq.~\eqref{eq:m_eff}) imposed by different prior choices for $\tau$, when the total number of input variables is $D=10$ and $D=1000$. $\tau_0$ is computed from formula~\eqref{eq:tau0} assuming $n=100$ observations with $\sigma=1$ and $p_0 = 5$ as the prior guess for the number of relevant variables. Note the varying scales on the horizontal axes in the bottom row plots.}
	\label{fig:meff_prior}
\end{figure*}

Figure~\ref{fig:meff_prior} shows histograms of prior draws for $m_\tx{eff}$ for some different prior choices for $\tau$, with total number of inputs $D=10$ and $D=1000$, assuming $n=100$ observations with $\sigma = 1$.
The first three priors utilize the value $\tau_0$ which is computed from~\eqref{eq:tau0} using $p_0 = 5$ as our hypothetical prior guess for the number of relevant variables.
Fixing $\tau=\tau_0$ results in a nearly symmetric distribution around $p_0$, while a half-normal prior with scale $\tau_0$ yields a skewed distribution favoring solutions with $m_\tx{eff} < p_0$ but allowing larger value to also be accommodated.
The half-Cauchy prior behaves similarly to the half-normal, but results in a distribution with a much thicker tail giving substantial mass also to values much larger than $p_0$ when $D$ is large. 
Figure~\ref{fig:meff_prior} also illustrates why the prior $\tau \sim \halfCauchy{0,1}$ is often a dubious choice: it places far too much mass on large values of $\tau$, consequently favoring solutions with most of the coefficients unshrunk.
Thus when only a small number of the variables are relevant -- as we typically assume -- this prior results in sensible inference only when $\tau$ is strongly identified by data.
Note also that, if we changed the value of $\sigma$ or $n$, the first three priors for $\tau$ would still impose the same prior for $m_\tx{eff}$, but this is not true for $\tau\sim \halfCauchy{0,1}$.

This way, by studying the prior for $m_\tx{eff}$, one can easily choose the prior for  $\tau$ based on the beliefs about the number of nonzero parameters.
Because the prior beliefs  can vary substantially for different problems and the results depend on the information carried by the data, there is no globally optimal prior for $\tau$ that works for every single problem.
Some recommendations, however, will be given in Section~\ref{sec:conclusion} based on these theoretical considerations and experiments presented in Section~\ref{sec:experiments}.
 
We shall conclude this section by pointing out a connection between our reference value~\eqref{eq:tau0} and the oracle result~\eqref{eq:vanderpas} for the simplified model~\eqref{eq:model_simple}. 
As pointed out in the last section, model~\eqref{eq:model_simple} corresponds to setting $\vc X = \vc I$ (which implies $n=D$ and $\vc X^\tp \vc X = \vc I$) in the usual regression model~\eqref{eq:lgm}.
Using this fact and repeating the steps needed to arrive at~\eqref{eq:tau0}, we get 
\begin{align}
	\tau_0 = \frac{p_0}{D-p_0} \sigma.
\label{eq:tau0_simple}
\end{align}
Suppose now that we select $p_0 = p^*$, that is, our prior guess is oracle.
Using the same assumptions as \cite{vanDerPas2014}, namely that $n,p^* \rightarrow \infty$ and $p^* = o(n)$, and additionally that $\sigma = 1$, we get $\tau_0 \rightarrow p^* / D = \tau^*$. This result is natural, as it means it is optimal to choose $\tau$ so that the imposed prior for the effective number of nonzero coefficients $m_\tx{eff}$ is centered at the true number of nonzeros $p^*$. 
This further motivates why $m_\tx{eff}$ is a useful quantity.

\subsection{Non-Gaussian Likelihood}
\label{sec:nongaussian_lik}

When the observation model is non-Gaussian, the exact analysis from Section~\ref{sec:meff_and_tau} is analytically intractable.
We can, however, perform the analysis using a Gaussian approximation to the likelihood.
Using the second order Taylor expansion for the log likelihood, the approximate posterior for the regression coefficients given the hyperparameters becomes (see the supplementary material for details)
\begin{align*}
	p(\vs \beta \given \vs \Lambda, \tau,\phi,\mc D) &\approx \Normal{\vs \beta \given \vs {\bar \beta}, \vs \Sigma}, \\
	\vs {\bar \beta} &= \tau^2 \vs \Lambda \left( \tau^2 \vs \Lambda + (\vc X^\tp \vs {\ti \Sigma}^{-1} \vc X)^{-1} \right)^{-1} \vs {\hat \beta}, \\
	\vs \Sigma &= (\tau^{-2}\vs \Lambda^{-1} + \vc X^\tp \vs {\ti \Sigma}^{-1} \vc X)^{-1},
\end{align*}
where $\vc {\ti z} = (\ti z_1,\dots,\ti z_n)$, $\vs {\ti \Sigma} = \tx{diag}(\ti \sigma_1^2,\dots,\ti \sigma_n^2)$ and $\vs {\hat \beta} = (\vc X^\tp \vs {\ti \Sigma}^{-1} \vc X)^{-1} \vc X^\tp \vs {\ti \Sigma}^{-1} \vc {\ti z}$ (assuming the first inverse  exists).
Here $\phi$ denotes the possible dispersion parameter and $(\ti z_i,\ti \sigma_i^2)$ the location and variance for the $i$th Gaussian pseudo-observation.
The fact that some of the observations are more informative than others ($\ti \sigma_i^2$ vary) makes  further simplification somewhat difficult.

To proceed, we make the rough assumption that we can replace each $\ti \sigma_i^2$ by a single variance term $\ti \sigma^2$.
Assuming further that the covariates are uncorrelated with zero mean and unit variance (as in Sec.~\ref{sec:meff_and_tau}), the posterior mean for the $j$th coefficient satisfies $\bar \beta_j = (1-\kappa_j)\hat \beta_j$ with shrinkage factor given by
\begin{align}
	\kappa_j = \frac{1}{1 + n \ti \sigma^{-2}\tau^2\lambda_j^2}.
\label{eq:shrinkage_factor_nongaussian}
\end{align}
The discussion in Section~\ref{sec:meff_and_tau} therefore also approximately holds for the non-Gaussian observation model, except that $\sigma^2$ is replaced by $\ti \sigma^2$.
Still, this leaves us with the question, which value to choose for $\ti \sigma^2$ when using this result in practice?

\begin{figure}[t]
\centering
	\setlength{\figureheight}{0.35\textwidth}%
	\setlength{\figurewidth}{0.45\textwidth}%
	\pgfplotsset{
	compat=newest,
	scaled ticks=false, tick label style={/pgf/number format/fixed},
	major tick length={0.03cm},
	x tick label style={font=\tiny},
	y tick label style={font=\tiny},
	xlabel style={font=\footnotesize, yshift=0.2em},
	ylabel style={font=\footnotesize},
	legend style={font=\tiny}, 
	} 
	\minput[pdf]{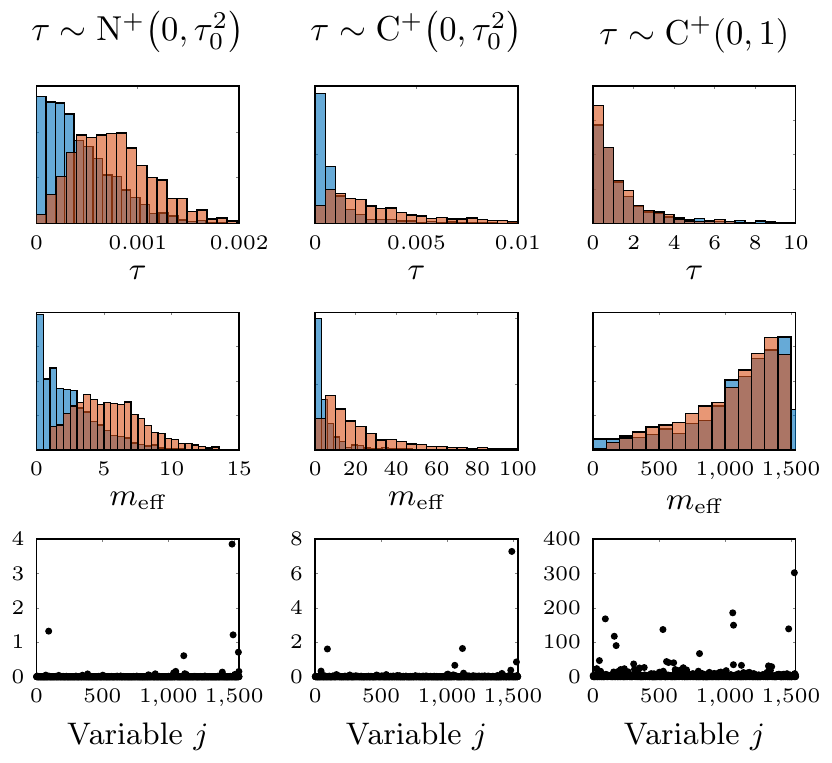}
	\caption{{\it Ovarian dataset}\,: Histograms of prior (blue) and posterior (red) samples for $\tau$ (top row) and $m_\tx{eff}$ (middle row), and absolute values of the posterior means for the regression coefficients $|\bar\beta_j|$ (bottom row) imposed by three different choices of prior for $\tau$. $\tau_0$ is computed from~\eqref{eq:tau0} using $p_0=3$ and $\sigma=2$ (for reasons explained in Sec.~\ref{sec:nongaussian_lik}). Note the changing scales on the horizontal axes in the top and middle row plots.}
	\label{fig:ovarian}
\end{figure}

We consider binary classification here as an example.
It can be shown (see the supplementary material) that for the logistic regression
\begin{align*}
	p(y_i=1\given f_i) = s(f_i) = \frac{1}{1+\exp(-f_i)}, \quad f_i = \vs \beta^\tp \vc x_i,
\end{align*}
for those points that lie on the classification boundary, we have $\ti \sigma_i^2=4$, and for others $\ti \sigma_i^2 > 4$.
For this reason we propose to use the results of Section~\ref{sec:meff_and_tau} as they are, by plugging in $\sigma=2$.
In practice this introduces some error, but the good thing is that we know in which way the results are biased.
For instance, because the true (unknown) effective noise deviation would be $\sigma>2$, the prior mean~\eqref{eq:meff_mean} using $\sigma=2$ is actually an {\it overestimate} of the true value.
Thus also when using the result~\eqref{eq:tau0}, we tend to favor slightly too small values of $\tau$ and thus also solutions with slightly less than $p_0$ nonzero coefficients.
In practice we observe that this approach, though relying on several crude approximations, is still useful and gives reasonably accurate results.

Finally, we note that similar approximate  substitute values for $\sigma$ to be used in equations of Section~\ref{sec:meff_and_tau} could also be derived for other link functions and observation models, but due to limited space, we do not focus on them in this paper.

\subsection{Other Shrinkage Priors}

Our approach could also be used with shrinkage priors other than the horseshoe,  as long as the prior can be written as scale mixtures of Gaussians like~\eqref{eq:hs} with some prior for the local scales $\lambda_j \sim p(\lambda_j)$.
An example of such is the hierarchical shrinkage prior -- a computationally more robust generalization of the horseshoe obtained by replacing the half-Cauchy priors in~\eqref{eq:hs} by half-$t$ distributions with some small degrees of freedom \citep{piironen2015}.
However, the closed form equations in Section~\ref{sec:meff_and_tau} apply only to the horseshoe.
Depending on the choice of $p(\lambda_j)$, corresponding analytical results may or may not be available, but as long as one is able to sample both from $p(\tau)$ and $p(\lambda_j)$, it is always easy to draw samples from the prior distribution for $m_\tx{eff}$.

\begin{figure*}[t]
\centering
	\setlength{\figureheight}{0.4\textwidth}
	\setlength{\figurewidth}{0.92\textwidth}
	\pgfplotsset{
	compat=newest,
	major tick length={0.07cm},
	x tick label style={font=\tiny},
	y tick label style={font=\tiny},
	xlabel style={font=\footnotesize},
	ylabel style={font=\footnotesize},
	legend style={font=\tiny},
	} 
	\minput[pdf]{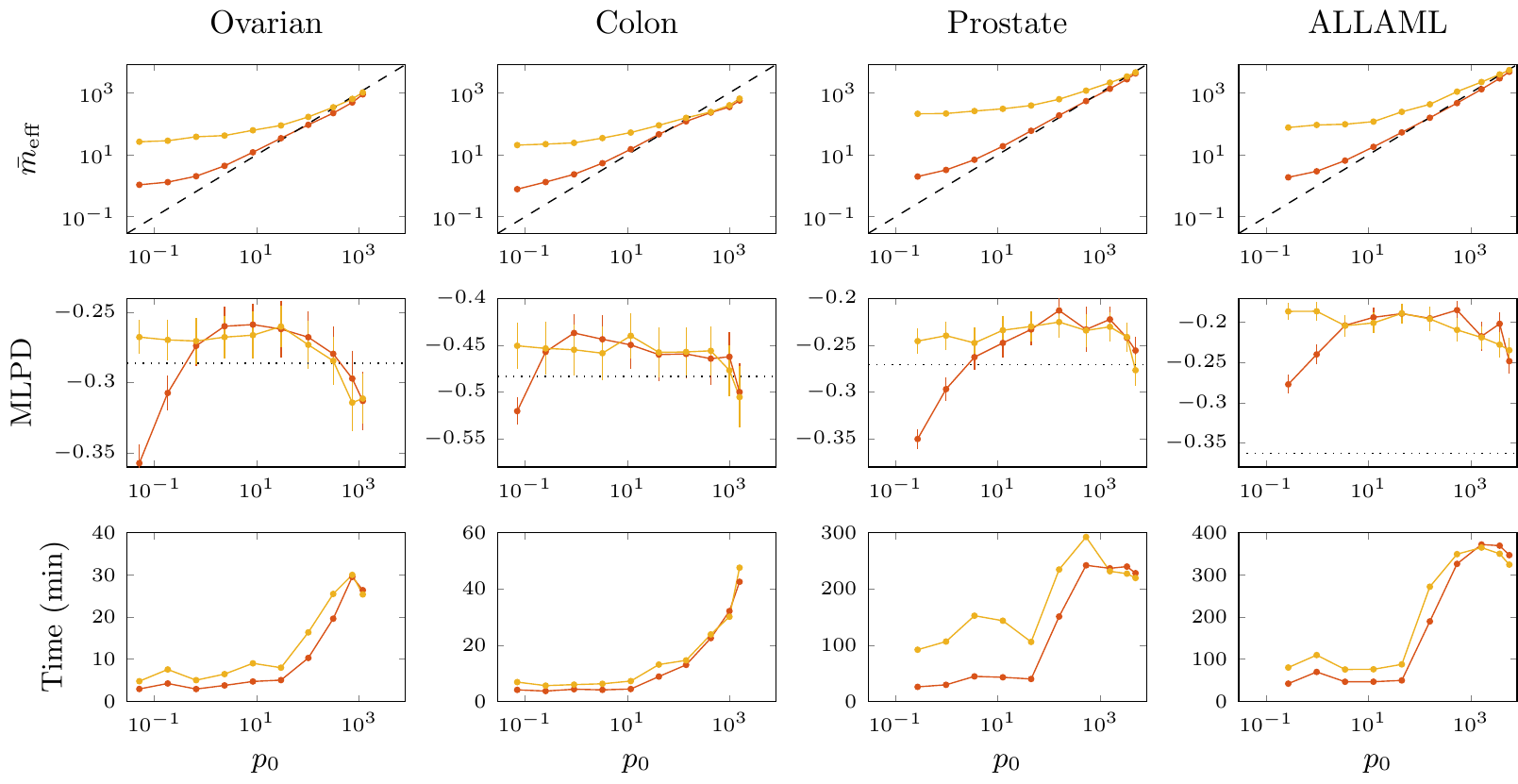}
	\caption{{\it Classification datasets}\,: Posterior mean for $m_\tx{eff}$, mean log predictive density (MLPD) on test data ($\pm$ one standard error), and computation time for two priors for the global hyperparameter:  $\tau \sim \halfNormal{0,\tau_0^2}$ (red), and $\tau \sim \halfCauchy{0,\tau_0^2}$ (yellow), where $\tau_0$ is computed from~\eqref{eq:tau0} varying $p_0$ (horizontal axis). For each curve, the largest $p_0$ corresponds to $\tau_0 = 1$. Dotted line in the middle row plots denotes the MLPD for LASSO. All the results are averaged over 50 random splits into training and test sets.}
	\label{fig:results_classif}
\end{figure*}

\begin{figure*}[t]
\centering
	\setlength{\figureheight}{0.4\textwidth}
	\setlength{\figurewidth}{0.92\textwidth}
	\pgfplotsset{
	compat=newest,
	major tick length={0.07cm},
	x tick label style={font=\tiny},
	y tick label style={font=\tiny},
	xlabel style={font=\footnotesize},
	ylabel style={font=\footnotesize},
	legend style={font=\tiny},
	} 
	\minput[pdf]{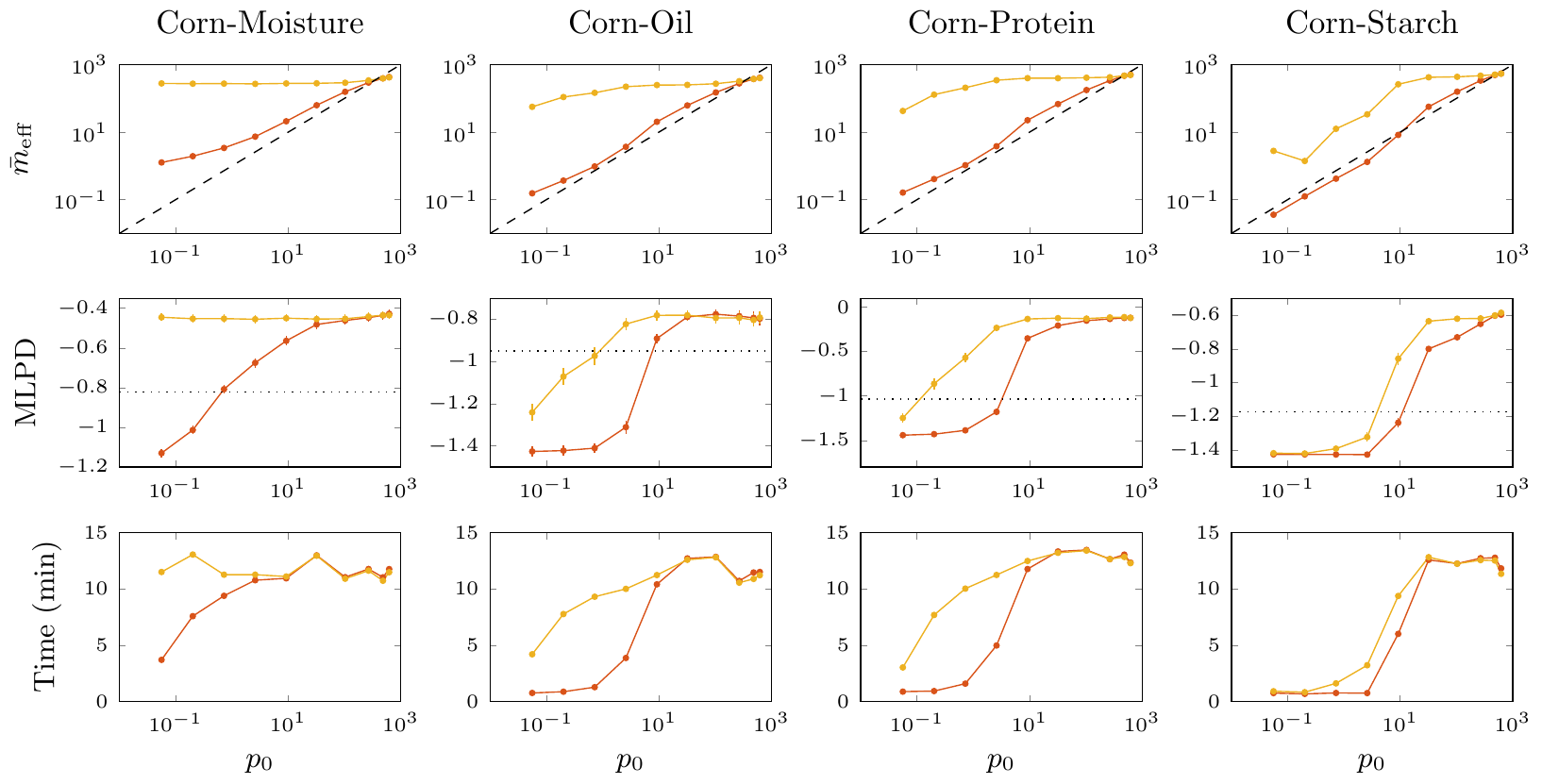}
	\caption{{\it Regression datasets}\,: Same as in Figure~\ref{fig:results_classif} but for the regression datasets. For each curve, the largest $p_0$ corresponds to $\tau_0 = \sigma$. }
	\label{fig:results_reg}
\end{figure*}

\section{EXPERIMENTS}
\label{sec:experiments}

We illustrate the importance of the prior choice for $\tau$ with real world examples (see the supplementary material for an additional experiment on synthetic data).
The datasets are summarized in Table~\ref{tab:datasets} and can be found online.\footnote{Colon, Prostate and ALLAML: \url{http://featureselection.asu.edu/datasets.php}; Ovarian data: request from the first author if needed; Corn data: \url{http://software.eigenvector.com/Data/Corn/
index.html}}
The first four are microarray cancer datasets, some of which have been used as benchmark datasets by several authors \citep[see e.g.,][and references therein]{hernandezlobato2010}.
The Corn data \citep{chen2009} consists of three sets of predictors (`m5', `mp5', `mp6') and four responses.
We use the `mp5' input data and consider prediction for all four targets.

\begin{table}[t]%
\centering
\abovetopsep=2pt
\caption{Summary of the datasets, number of predictor variables $D$ and dataset size $n$. Classification datasets are all binary.}
\label{tab:datasets}
\begin{tabular}{ llccc }
\toprule
Dataset & Type & $D$ & $n$ \\ 
\midrule
Ovarian & Classification & 1536 & 54   \\
Colon & Classification & 2000 & 62   \\
Prostate & Classification & 5966 & 102   \\
ALLAML & Classification & 7129 & 72   \\
Corn (4 targets) & Regression & 700 & 80   \\ 
\bottomrule
\end{tabular}
\end{table}

A Gaussian linear model was used for the regression tasks and logistic regression for classification.
The horseshoe prior was employed for the regression coefficients and a weakly informative prior $\beta_0 \sim \Normal{0,10^2}$ for the intercept.
The noise variance was given the standard prior $p(\sigma^2)\propto 1/\sigma^{2}$ in the regression problems.
All considered models were fitted using Stan (version~2.12.0, codes in the supplementary material), running 4 chains, 1000 samples each, first halves discarded as warmup.

{\it Effect on parameter estimates}\quad
We first consider the Ovarian dataset as a representative example of how much the prior choice $p(\tau)$ can affect the parameter estimates. 
We fitted the model to the data using three different priors for the global parameter; $\tau \sim \halfNormal{0,\tau_0^2}$, $\tau \sim \halfCauchy{0,\tau_0^2}$, and $\tau \sim \halfCauchy{0,1}$, where $\tau_0$ is computed from~\eqref{eq:tau0} using $p_0 = 3$ as our prior guess for the number of relevant variables and $\sigma = 2$ (for reasons discussed in Section~\ref{sec:nongaussian_lik}).

Figure~\ref{fig:ovarian} shows prior and posterior samples for $\tau$ and $m_\tx{eff}$, and the absolute values of the posterior means for the regression coefficients, for the three prior choices.
The results for $\tau \sim \halfCauchy{0,1}$ illustrate how weakly $\tau$ is identified by the data: there is very little difference between the prior and posterior samples for $\tau$ and consequently for $m_\tx{eff}$, and thus this ``non-informative'' prior actually has a strong influence on the posterior inference.
The prior results in a severe under-regularization and implausibly large absolute values for the logistic regression coefficients (magnitude in the hundreds).

Replacing the scale of the half-Cauchy with $\tau_0$, reflecting a more sensible guess for the number of relevant variables, has a substantial effect on the posterior inference: the posterior mass for $m_\tx{eff}$ becomes concentrated on more reasonable values and the magnitude of the regression coefficients more sensible.
Replacing the half-Cauchy by half-normal places a tighter constraint on $\tau$ and consequently also on $m_\tx{eff}$.
Either of these two choices is a substantial improvement over $\tau \sim \halfCauchy{0,1}$, and this is also seen in the predictive accuracy (to be discussed in a moment).

A potential explanation for why $\tau$ and therefore $m_\tx{eff}$ are not strongly identified here is that there are a lot of correlations in the data.
For instance, the predictor $j=1491$ which appears relevant based on its regression coefficient, has an absolute correlation of at least $|\rho|=0.5$ with 314 other variables, out of which 65 correlations exceed $|\rho|=0.7$.
This indicates that there are a lot of potentially relevant but redundant predictors in the data.
Note also that even though our theoretical treatment for $m_\tx{eff}$ assumes uncorrelated variables, the correlations do not seem to be a marked problem in this sense.

{\it Prediction accuracy and computation time}\quad
To investigate the effect of the prior choice $p(\tau)$ on the prediction accuracy, we splitted each dataset into two halves, using one fifth of the data as a test set.
All the results were then averaged over 50 such random splits into training and test sets.
For the regression problems, we carried out the tests for priors $\tau \given \sigma \sim \halfNormal{0,\tau_0^2}$ and $\tau \given \sigma \sim \halfCauchy{0,\tau_0^2}$ with $\tau_0$ given by Eq.~\eqref{eq:tau0} for various $p_0$.
The experiments for the classification datasets were done using the same priors by plugging in $\sigma=2$.
We also compared the prediction accuracies to LASSO with the regularization parameter tuned by 10-fold cross-validation, and noise variance in regression computed by dividing the sum of squared residuals by $n-p_\tx{lasso}$, where $p_\tx{lasso}$ is the size of the LASSO active set \citep{reid2016}.

Figures~\ref{fig:results_classif}~and~\ref{fig:results_reg} show the effect of the prior choice on the posterior mean for $m_\tx{eff}$, test prediction accuracy, and computation time (wall time).
The classification datasets illustrate a clear benefit from using even a crude prior guess for the number of relevant variables $p_0$: there is a relatively wide range of values for $p_0$ which yield simpler models (smaller $\bar m_\tx{eff}$), better predictive accuracy, and substantially reduced computation time when compared to using $\tau_0 = 1$ (the largest $p_0$ for each curve), regardless of whether half-normal or half-Cauchy prior is used.
The reduced computation time is due to tighter concentration of the posterior mass which aids the sampling greatly.
The half-Cauchy prior performs better than the half-normal prior with small $p_0$ values (both for classification and regression datasets), because the thick tails allow $\tau$ to get much larger values than $\tau_0$, but for too small values of $p_0$, even the half-Cauchy can lead to bad results (Corn-Oil -- Corn-Starch). 
On the other hand, when $p_0$ happens to be well chosen, the tighter half-normal prior can yield a simpler model (smaller $m_\tx{eff}$) with equally good or even better predictive accuracy (classification datasets).

For any reasonably selected prior, the horseshoe consistently outperforms LASSO in a pairwise comparison, but note that for Ovarian and Colon, $\tau_0 = 1$ (largest $p_0$) actually yields worse results.
In terms of classification accuracy, the differences are smaller, but in regression, the horseshoe also performs better when measured by the mean squared error (not shown). 
A clear advantage for LASSO, on the other hand, is that it is hugely faster, with computation time of less than a second for these problems.

\section{CONCLUSION AND RECOMMENDATIONS}
\label{sec:conclusion}

This paper has discussed the prior choice for the global shrinkage parameter in the horseshoe prior for sparse Bayesian regression and classification.
We have shown that the previous default choices are often dubious based on their tendency to favor solutions with too many parameters unshrunk.
The experiments show that for many datasets, one can obtain clear improvements -- in terms of better parameter estimates, prediction accuracy and computation time -- by coming up even with a crude guess for the number of relevant variables and transforming this knowledge into a prior for $\tau$ using our proposed framework.
The results also show that there is no globally optimal prior choice that would perform best for all problems, which emphasizes the relevance of the prior choice.

As a new default choice for regression, we recommend $\tau \given \sigma \sim \halfCauchy{0,\tau_0^2}$, where $\tau_0$ is computed from~\eqref{eq:tau0} using the prior guess $p_0$ for the number of relevant variables.
For logistic regression, an approximately equivalent choice is obtained by plugging in $\sigma=2$ (see Sec.~\ref{sec:nongaussian_lik}).
This choice seems to perform well unless $p_0$ is very far from the optimal.
If the results still indicate that more regularization is needed, we recommend investigating the imposed prior on $m_\tx{eff}$ as discussed in Section~\ref{sec:meff_and_tau}, and changing $p(\tau)$ so that the prior for $m_\tx{eff}$ corresponds to our beliefs as closely as possible.

\subsubsection*{Acknowledgements}

We thank Andrew Gelman, Jonah Gabry and Eero Siivola for helpful comments to improve the manuscript.
We also acknowledge the computational resources provided by the Aalto Science-IT project.

\newpage
{
\renewcommand{\section}[2]{}%
\subsubsection*{References}
\bibliographystyle{apalike}
\bibliography{references}
}

\newpage
\section*{SUPPLEMENTARY MATERIAL}

\subsection*{Non-Gaussian Likelihood}

{\it Posterior approximation}\quad Using the second order Taylor expansion for the log likelihood terms $L_i(f_i,\phi)$ (where $f_i = \vs \beta^\tp \vc x_i$ and $\phi$ denotes a possible dispersion parameter), we can approximate the posterior as \citep[ch.~16.2,][]{gelman2013book}
\begin{align*}
	&\log p(\vs \beta \given \vs \Lambda, \tau,\phi,\mc D) \\
	&\approx \log p(\vs \beta \given \vs \Lambda, \tau,\phi) - \sum_{i=1}^n \frac{1}{2\ti \sigma_i^2}(\ti z_i - f_i)^2 + \tx{const.},
\end{align*}
where 
\begin{align*}
	\ti z_i &= f_i - \frac{L_i'(f_i,\phi)} {L_i''(f_i,\phi)}, \quad
	\ti \sigma^2_i = - \frac{1} {L_i''(f_i,\phi)},
\end{align*}
denote the location and variance of the Gaussian pseudo-observations.
The derivatives are calculated w.r.t. $f_i$ at the posterior mode $\bar f_i = \vs {\bar \beta}^\tp \vc x_i$.
Using these, the posterior (given the hyperparameters) is approximately
\begin{align*}
	p(\vs \beta \given \vs \Lambda, \tau,\phi,\mc D) &\approx \Normal{\vs \beta \given \vs {\bar \beta}, \vs \Sigma}, \\
	\vs {\bar \beta} &= \tau^2 \vs \Lambda \left( \tau^2 \vs \Lambda + (\vc X^\tp \vs {\ti \Sigma}^{-1} \vc X)^{-1} \right)^{-1} \vs {\hat \beta}, \\
	\vs \Sigma &= (\tau^{-2}\vs \Lambda^{-1} + \vc X^\tp \vs {\ti \Sigma}^{-1} \vc X)^{-1},
\end{align*}
where $\vc {\ti z} = (\ti z_1,\dots,\ti z_n)$, $\vs {\ti \Sigma} = \tx{diag}(\ti \sigma_1^2,\dots,\ti \sigma_n^2)$ and $\vs {\hat \beta} = (\vc X^\tp \vs {\ti \Sigma}^{-1} \vc X)^{-1} \vc X^\tp \vs {\ti \Sigma}^{-1} \vc {\ti z}$ (assuming the first inverse  exists).

\vspace{0.7cm}
\textit{Logistic regression}\quad
Consider the logistic regression model
\begin{align*}
	p(y_i=1\given f_i) = s(f_i) = \frac{1}{1+\exp(-f_i)}.
\end{align*}
The second derivative for the $i$th log-likelihood term is given by
\begin{align*}
	L_i''(f_i) &= \left(\frac{y_i}{s(f_i)} - \frac{1-y_i}{1-s(f_i)} \right) s''(f_i) \\
					&- \left(\frac{y_i}{s(f_i)^2} + \frac{1-y_i}{(1-s(f_i))^2} \right) (s'(f_i))^2.
\end{align*}
If we now plug in the derivatives 
\begin{align*}
	s'(f_i) &= s(f_i)(1-s(f_i)), \\
	s''(f_i) &= s'(f_i)(1-2s(f_i)),
\end{align*}
after a few lines of straightforward algebra, we are left with
\begin{align*}
	L_i''(f_i) = s(f_i)(s(f_i)-1).
\end{align*}
This is a strictly negative function with minimum at $s(f_i)=\frac{1}{2}$, which occurs when $f_i = 0$.
Thus also $\ti \sigma_i^2 = -1/L_i''(f_i)$ is minimized at $f_i = 0$.
In other words, those points that lie on the classification boundary are the most informative ones, and the pseudo-variance for these points is
\begin{align*}
	\ti \sigma_i^2 = -\frac{1}{\frac{1}{2} \left(-\frac{1}{2}\right)} = 4.
\end{align*}
This result serves as a useful reference value as discussed in Section~\ref{sec:nongaussian_lik}.

\subsection*{Additional experiments}

\begin{figure}[t]
\centering
	\setlength{\figureheight}{0.18\textwidth}
	\setlength{\figurewidth}{0.45\textwidth}
	\pgfplotsset{
	compat=newest,
	major tick length={0.07cm},
	x tick label style={font=\tiny},
	y tick label style={font=\tiny},
	xlabel style={font=\footnotesize},
	ylabel style={font=\footnotesize},
	legend style={font=\tiny},
	}
	\minput[pdf]{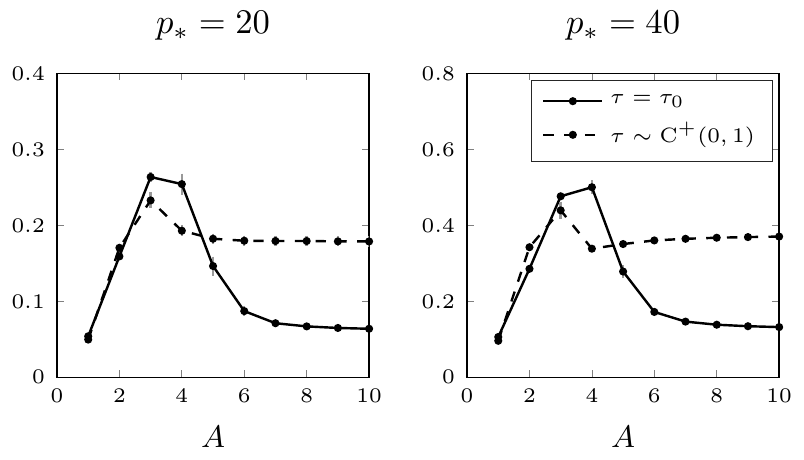}
	\caption{{\it Synthetic example}\,: Mean squared error (MSE) between the estimated and the true coefficient vector of length $n=400$ on average over 100 different data realizations. The true coefficient vector has either $p_*=20$ or $p_*=40$ elements with a nonzero value equal to $A$ and the rest of the coefficients are set to zero.}
	\label{fig:synthetic1} 
\end{figure}

\begin{figure*}[t]
\centering
	\setlength{\figureheight}{0.35\textwidth}
	\setlength{\figurewidth}{0.85\textwidth}
	\pgfplotsset{
	compat=newest,
	major tick length={0.07cm},
	x tick label style={font=\tiny},
	y tick label style={font=\tiny},
	xlabel style={font=\footnotesize},
	ylabel style={font=\footnotesize}, 
	ylabel style={text width=5em, rotate=-90, align=left},
	legend style={font=\tiny},
	} 
	\minput[pdf]{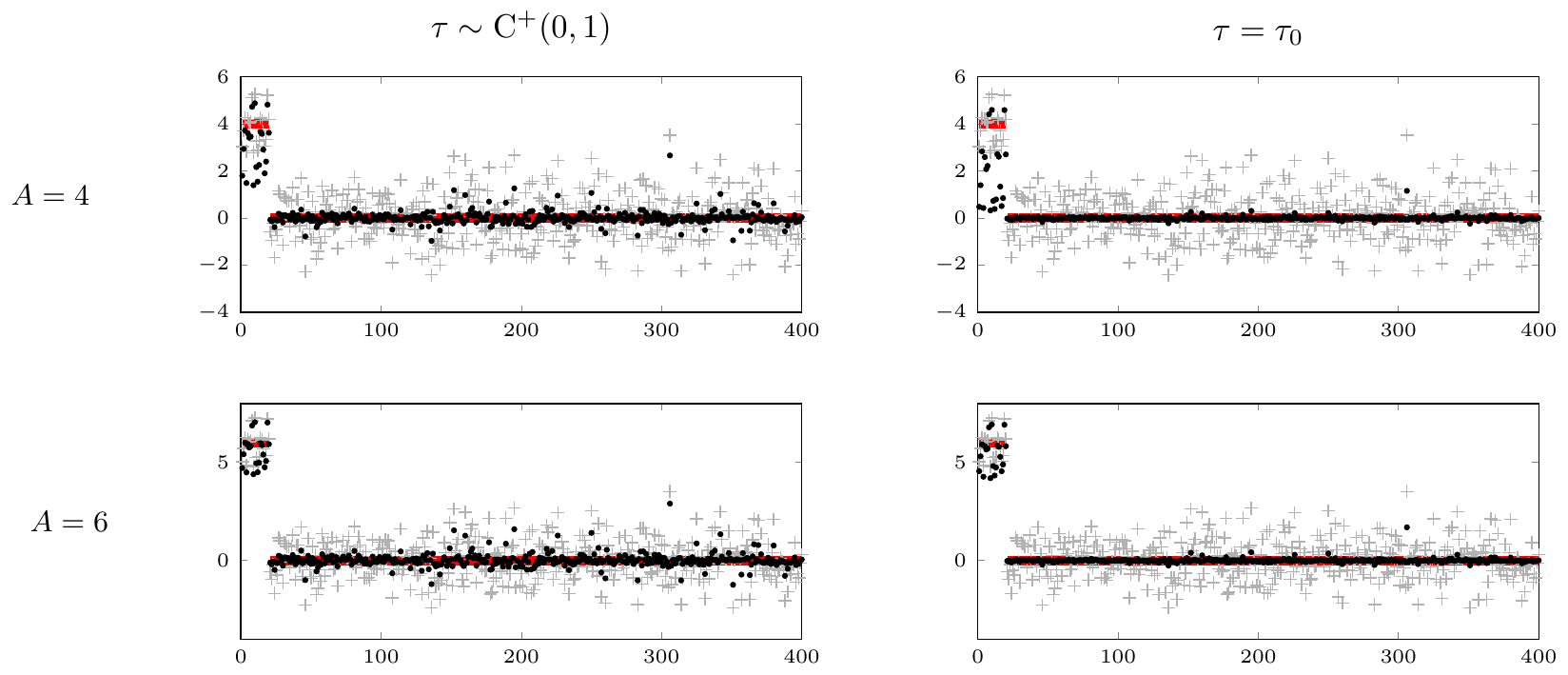}
	\caption{{\it Synthetic example}\,: An example data realization $\vc y$ (crosses), posterior mean $\vs{\bar \beta}$ (dots) and the true signal $\vs \beta_*$ (red lines) for $A=4$ and $A=6$ when $p_*=20$. In both cases the oracle value for $\tau$ helps to shrink the zero components in $\vs \beta$ but also overshrinks the actual signals in the case $A=4$.}
	\label{fig:synthetic2} 
\end{figure*}

We present here an additional synthetic experiment that could not fit to the main content of the paper.
The example is taken from \cite{vanDerPas2014}.
Consider model~\eqref{eq:model_simple}, where each $y_i$ is generated by adding Gaussian noise with $\sigma^2=1$ to the corresponding signal~$\beta_i$.
We generated 100 data realizations with $n=400$ and the true $\vs \beta_*$ having either $p_*=20$ or $p_*=40$ nonzero entries equal to $A=1,2,\dots,10$ with the rest of the entries being zeros.
We then computed the mean squared error (MSE) between the estimated posterior mean $\vs{ \bar \beta}$ and the true $\vs \beta_*$ for the prior $\tau\sim \halfCauchy{0,1}$ and for $\tau=\tau_0$, where $\tau_0$ is computed from Equation~\eqref{eq:tau0_simple} with the oracle prior guess $p_0=p_*$.
The purpose of this setup is to further demonstrate how one could benefit from the prior knowledge about the sparsity of $\vs \beta$ using our framework, provided such prior knowledge exists.
Notice though, that also in the latter case $\tau$ has a distribution because it depends on~$\sigma$ which is treated as an unknown parameter.

Figure~\ref{fig:synthetic1} shows the MSE for the two priors for the different values of $p_*$ and $A$.
For both priors the error is largest around $A\approx 3.5$, which is called the ``universal threshold" by \cite{vanDerPas2014}.
Below this threshold the nonzero components in $\vs \beta$ are too small to be detected and are thus shrunk too heavily towards zero which introduces error.
For $A=4$ the oracle prior actually yields worse results due to this overshrinkage (see discussion below), but gives clearly superior results for larger $A$.

Figure~\ref{fig:synthetic2} illustrates the data $\vc y$ and the estimated coefficients $\vs{\bar \beta}$ for one particular data realization when $A=4$ and $A=6$.
In both cases the oracle choice of $\tau$ helps to shrink the zero components in $\vs \beta$ towards zero, but for $A=4$ also overshrinks the nonzero components. 
The reason for the overshrinkage is that some observations $y_i$ that correspond to zero signal ($\beta_i=0$) happen to have similar magnitude to the observations coming from an actual signal ($\beta_i=A$), and thus these irrelevant components ``steal" from the limited budget for $m_\tx{eff}$.
For this particular value of $A$ (and $p_*$) the overshrinkage of the actual signals happens to be worse in terms of MSE than undershrinkage of the zero components, and thus one would get better results by setting $p_0$ to be slightly above the true $p_*$ (results not shown).
For $A=6$ the actual signals are large enough to be distinguished from zero, and the oracle selection of $\tau$ yields substantially better estimate for $\vs \beta$.

\clearpage
\onecolumn
\subsection*{Stan codes}

The following shows the Stan code for the linear Gaussian model.
We use the parametrization proposed by \cite{peltola2014} (codes at \url{https://github.com/to-mi/stan-survival-shrinkage}) as it is more robust for sampling than the literal~\eqref{eq:hs}.
Even with this parametrization we usually set {\tt adapt\_delta} $=0.99$ when calling Stan, as this can sometimes reduce the number of divergent transitions which can be an issue for the horseshoe prior \citep[see][]{piironen2015}.

In the code, both $\tau$ and $\lambda_j$ are given half-$t$ priors with the degrees of freedom and the scale defined by the user (the scale can be adjusted only for $\tau$, the local parameters $\lambda_j$ have unit scale).
Setting {\tt nu\_local} $=1$ corresponds to the horseshoe.
{\tt nu\_global} $ = 1$ gives $\tau$ a half-Cauchy prior, whereas fixing {\tt nu\_global} to some large value (say 100) would give $\tau$ practically a half-normal prior.
The scale for $\tau$ is {\tt scale\_global*sigma}, so if we want to set this to be $\tau_0=\frac{p_0}{D-p_0} \frac{\sigma}{\sqrt{n}}$ (Eq.~\eqref{eq:tau0}), we should set \texttt{scale\_global} $ = \frac{p_0}{(D-p_0)\sqrt{n}}$. 

\lstset{language=C, tabsize=4, basicstyle=\ttfamily\scriptsize, literate={~} {$\sim$}{1}} 
\lstinputlisting{\codesdir lm_gaussian_hs.stan}

\newpage
The code for the logistic regression model is very similar, we simply remove the lines related to the noise deviation {\tt sigma}, and change the observation model and the type of the target variable data {\tt y}.
The scale for $\tau$ is now simply {\tt scale\_global}.
Thus, to follow our recommendation, we set {\tt scale\_global} $=\tau_0=\frac{p_0}{D-p_0} \frac{\sigma}{\sqrt{n}}$ (Eq.~\eqref{eq:tau0}), by plugging in $\sigma= 2$ (Sec.~\ref{sec:nongaussian_lik}).

\lstset{language=C, tabsize=4, basicstyle=\ttfamily\scriptsize, literate={~} {$\sim$}{1}}
\lstinputlisting{\codesdir glm_logit_hs.stan}

\end{document}